\documentclass[aps,floats,floatfix,amssymb,amsmath,prb,twocolumn,superscriptaddress]{revtex4-2}

\usepackage{graphicx}
\usepackage[colorlinks=true,citecolor=blue,linkcolor=blue,urlcolor=blue,bookmarks=true]{hyperref}

\usepackage[utf8]{inputenc}
\usepackage{polski}
\usepackage[english]{babel}
\usepackage{amsmath}
\usepackage{xcolor}

\definecolor{nred} {RGB}{224,0,0}
\definecolor{nblue} {RGB}{28,130,185}
\definecolor{dgreen}{RGB}{78,138,21}
\definecolor{norange}{RGB}{230,120,20}

\DeclareUnicodeCharacter{2212}{-}
 
\begin{document}

\title{Finite-temperature properties of the easy-axis Heisenberg model on frustrated lattices}  
\author{M. Ulaga}
\affiliation{Jo\v zef Stefan Institute, SI-1000 Ljubljana, Slovenia}
\author{J.  Kokalj}
\affiliation{Faculty of Civil and Geodetic Engineering, University of Ljubljana, SI-1000 Ljubljana, Slovenia}
\affiliation{Jo\v zef Stefan Institute, SI-1000 Ljubljana, Slovenia}
\author{A. Wietek}
\affiliation{Max Planck Institute for the Physics of Complex Systems, Dresden 01187, Germany}
\author{A. Zorko}
\affiliation{Jo\v zef Stefan Institute, SI-1000 Ljubljana, Slovenia}
\affiliation{Faculty of Mathematics and Physics, University of Ljubljana, SI-1000 Ljubljana, Slovenia}
\author{P. Prelov\v{s}ek}
\affiliation{Jo\v zef Stefan Institute, SI-1000 Ljubljana, Slovenia}

\begin{abstract}
Motivated by recent experiments on a compound {displaying Ising-like short-range correlations on the triangular lattice,
we study the anisotropic easy-axis spin-$1/2$
Heisenberg model on the triangular and kagome lattice} by performing numerical calculations of
finite-temperature properties, in particular of static spin structure factor and of thermodynamic quantities,
on systems with up to 36 sites.  On the triangular lattice, the low-temperature spin structure factor {exhibits 
long-range} spin correlations in  the whole range of anisotropies, whereas thermodynamic quantities 
reveal  a crossover upon increasing the anisotropy, most pronounced in  the vanishing generalized Wilson 
ratio in the easy-axis regime.  In contrast, on the kagome lattice, the spin structure factor is short-range, and
thermodynamic quantities evolve steadily between the easy-axis and the isotropic case, consistent with the 
interpretation in terms of {a} spin liquid.

\end{abstract}

\maketitle

\section{Introduction}  
Quantum spin $S=1/2$ Heisenberg model (HM) on frustrated lattices  has  been attracting ongoing  theoretical interest 
ever since Anderson's seminal conjecture \cite{anderson73} that the HM with antiferromagnetic (AFM) 
exchange coupling between nearest neighbors on the triangular lattice (TL) can exhibit properties of quantum spin liquid (QSL).  
Theoretical studies intensified after the discovery of several classes of insulators with local magnetic 
moments \cite{mila00,lee08,balents10,savary17}, which do not reveal any magnetic 
long-range order (LRO) down to the lowest  experimentally accessible temperature $T$. 
The most established case of QSL ground state (gs) is the AFM HM on the kagome lattice (KL) 
even though the precise nature of the QSL is still under active debate \cite{mila98, budnik04,lauchli11,iqbal13,schnack18}.  
On the other hand, studies  of the isotropic HM on TL have revealed long-range order (LRO) at $T=0$ with
spins {in $120^\circ$ aligment} \cite{bernu94,capriotti99,white07,chernyshev09}. 

AFM HM on TL with anisotropic exchange has also been considered  theoretically since the 
ground-state (gs) properties  of the Ising limit have been evaluated  analytically, 
revealing finite remanent entropy $s_0 =0.323$  \cite{wannier50} as well as Curie-type susceptibility $\chi_0 \sim C/T$
at low $T$ \cite{sykes61,miyashita85,sano87}.  The extension including a weak transverse spin exchange
with relative $\alpha = J_\perp/J_z <1$ has been initially investigated in relation to possible stabilization
of QSL \cite{fazekas74,kleine92,kleine921} while more elaborate numerical studies revealed the persistence of 
gs long-range spin correlations \cite{wang09,jiang09,yamamoto14,sellmann15} in the whole range of anisotropies
$0 \le \alpha \leq 1$. The effect of quantum fluctuations on finite-$T$ properties has been so far mostly restricted 
to the analogous problem  of the frustrated Ising model  with an additional transverse field 
\cite{moessner00,moessner01, mostovoy03,chern08,chen19} while some numerical results of thermodynamic
quantities of anisotropic HM on modest-size frustrated lattices {have also been} performed \cite{isoda08,isoda11} 
to show the lifting of macroscopic degeneracy {by quantum} fluctuations introduced {via} $\alpha >0$. 

The motivation for the present study of finite-$T$ properties of anisotropic HM is the recent
discovery and study of the material neodymium heptatantalate (NdTa$_7$O$_{19}$) \cite{arh22}
with effective $S=1/2$ on a perfect TL, which  due to the strong spin-orbit coupling 
is expected to map on HM in the regime with strong easy-axis anisotropy,  but still with the crucial role of  quantum 
fluctuations. Inelastic neutron scattering revealed  Ising-like short-range spin correlations between nearest neighbors, while evidence 
of spin fluctuations persisting down to the lowest accessible $T$ was found via muon spectroscopy  \cite{arh22}
suggesting QSL behavior. 

In this paper, we present numerical results for thermodynamic quantities, including the entropy density $s(T)$,
specific heat $c(T)$, and the longitudinal magnetic susceptibility  $\chi_0(T)$, as well as the static spin structure  factor $S_{\bf q}(T)$. 
For comparison, we also discuss {these} quantities and their $T$-dependence within the anisotropic HM on the KL. 
In analogy with previous studies for the isotropic HM \cite{schnack18,prelovsek20},  we present results 
on lattices with  up to $N=36$ sites. It should be emphasised that due to large $s(T)$ at low $T$ in 
systems with $\alpha  \ll 1$, we are able to obtain reliable results even for very low $T$, i.e., typically
$T\gtrsim 0.1 \alpha J$. {The generalized Wilson ratio $R(T)$ has been used as a hallmark of possible QSL in the isotropic HM \cite{jaklic00,prelovsek18,prelovsek20,richter22}, 
expressing} the ratio of low-lying magnetic vs. all excitations.   In the anisotropic HM on TL, 
$R(T\to 0)$ reveals a qualitative change/crossover at $\alpha \sim 0.3$, i.e.,  from divergence at $\alpha =1$
to vanishing at $\alpha \gtrsim 0$,  indicating
that nonmagnetic $S^z=0$ gap $\Delta_0$ is well below  the magnetic gap which becomes finite with the 
departure from the Ising limit, i.e., $\Delta_1 \sim \alpha J/2$. On the other hand, spin correlations {$S_{{\bf q}}(T)$ at ${\bf q}_0$ in the corner of the Brillouin zone (BZ) still appear to diverge} at $T \to 0$, implying the persistence of gs 
LRO with rather modest dependence on  $\alpha$.  The easy-axis regime is accompanied also by {a} more pronounced
magnetization plateau at $m=1/3$ \cite{honecker04} at finite magnetic field $h$. 
Within the related HM on KL{,} the thermodynamic quantities behave  in a similar manner in the  regime of $\alpha \ll 1$, 
but in contrast to TL continuously evolve  into {the} isotropic QSL at $\alpha =1$, with vanishing $R(T \to 0)$. The essential
difference to TL is a large number of nonmagnetic excitations below the lowest magnetic excitation \cite{waldtmann98,
lauchli19,prelovsek20}, but also short-range spin correlations as manifested in $S_{\bf q}(T)$ in the whole range 
of $\alpha<1$. 

\section{Model and numerical method}
We consider the anisotropic $S=1/2$ HM with the nearest-neighbor exchange 
interaction $J$ {in} the presence of a longitudinal magnetic field
$h$,
\begin{equation}
	H= J  \sum_{\langle ij \rangle} \lbrack S^z_i S^z_j + \frac{\alpha}{2}( S^+_i S^-_j  + S^-_i S^+_j)  \rbrack
	+ \sum_{i} h S^z_i~~,~~\label{his}
\end{equation}
where the first sum runs over the nearest-neighbor pairs. We consider the easy-axis regime 
$\alpha \leq 1$ and we set $J=1$ as the unit of 
energy. We numerically study HM on the frustrated TL and KL with $N=18 - 36$ sites 
and periodic boundary conditions (PBC). 

We calculate thermodynamic quantities as well as $S_{\bf q}(T)$ by employing the  finite-temperature Lanczos method (FTLM) 
\cite{jaklic94,jaklic00}, used in numerous studies of $T>0$ properties of models of strongly correlated systems
\cite{prelovsek13}, including QSL models \cite{schnack18,prelovsek18,prelovsek20, prelovsek21}.  In the present study we
employ a highly parallelized code \cite{wietek18} and reach $N=36$ sites requiring
the handling of $N_{st} \sim 10^{10}$ basis states in the largest $S^z=0 $ sector.  To avoid the
considerable sampling $N_s >1$ over initial wavefunctions required by
FTLM, we use the orthogonal Lanczos method ~\cite{morita22} which treats the gs (within each sector) 
within the Lanczos procedure, and all other states  orthogonal to the gs in a standard FTLM approach, resulting in
considerably reduced number of required samples, i.e., $N_s \sim 3$.  

\begin{figure}[h!]
\centering
\includegraphics[width=0.9\columnwidth]{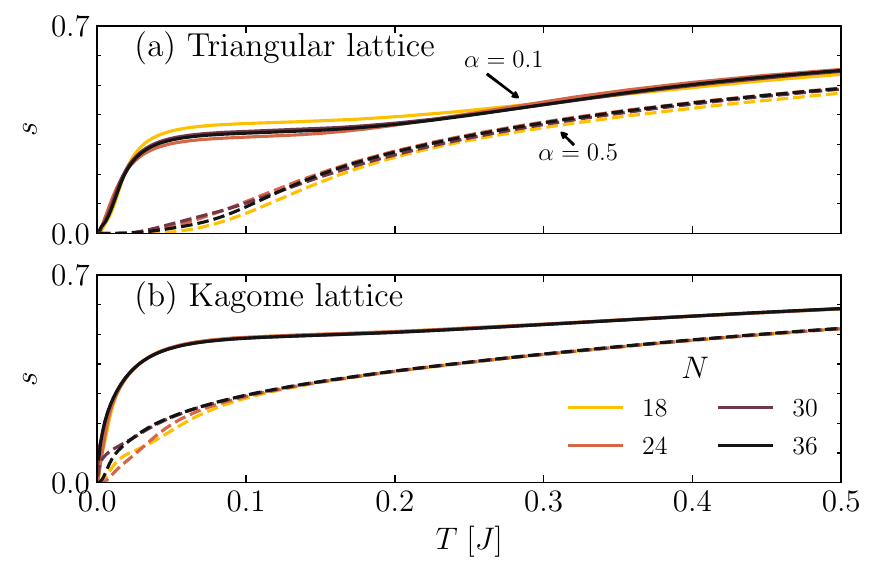}
\caption{ Comparison of FTLM results for the entropy $s(T)$, as obtained for different sizes $N = 18 - 36$
for two characteristic $\alpha = 0.1, 0.5$ (a) for the triangular lattice, and (b) for the kagome lattice.  }  
\label{figfs}
\end{figure}

First, we consider the $h=0$ case. The central quantity evaluated within FTLM for a given system 
is the grand-canonical sum $Z(T) = \textrm{Tr}\lbrace \exp[-(H-E_0)]/T\rbrace$ where $E_0$ is the gs energy. 
Orthogonalized FTLM reproduces exactly $Z(T \to 0)=1$ (for non-degenerate gs)  even for $N_{s}=1$. 
Within the same procedure, we evaluate the entropy density 
\begin{equation}
	s(T) = \lbrack \ln Z +( \langle H \rangle - E_0)/T \rbrack /N, \label{st}
\end{equation}
as well as the corresponding specific heat $c(T) = T (ds/dT)$ and the uniform (easy-axis) magnetic susceptibility 
$\chi_0 = {\cal M}^2/T$ (using theoretical units $k_B= g =\mu_h =1$) where the 
magnetization fluctuations are  ${\cal M}^2 = \langle (S^z)^2 \rangle /N $. 
Of special interest, in particular in relation to the QSL phenomenon, is the
generalized Wilson ratio \cite{jaklic00,prelovsek18,prelovsek20,richter22}
\begin{equation}
	R=  4 \pi^2 T \chi_0 / (3 s), \label{rw}
\end{equation}
which equals the standard Wilson ratio (constant at $T \to 0$) in the case of Fermi-liquid
behavior at  low $T$, i.e., for $s= c=\gamma T$. {It should be noted that a constant $R(T \to 0)=R_0$ 
appears also within the Ising limit ($\alpha=0$) since $\chi_0 \sim C/T$ and $s_0 >0$, so that 
$R_0=  4 \pi^2 C/ (3 s_0) > 0$.}  For the considered models at $\alpha >0$, this is not the case.
Still we have $R(T) \propto {\cal M}^2(T)/s(T)$ at $T>0$ which represents a measure for 
the ratio of easy-axis magnetic excitations (contained in ${\cal M}^2$)  to all excitations  (represented with 
$s$).  In particular, in the frustrated isotropic models, the signature of QSL is $R_0 \to 0$
\cite{prelovsek18,prelovsek20}, which is the case  for $\alpha=1$ model on KL, but not on TL, where the lowest 
magnetic excitation is a triplet leading to a diverging  $R_0 \to \infty$. 

It is relevant to realize the limitations of obtained numerical results for $T>0$.
With the use of orthogonalized FTLM, statistical fluctuations at fixed $N$ are suppressed 
even at $T \to 0$, so the actual limitations are finite-size effects.
For thermodynamic quantities, it is essential to capture enough many-body states. This requires
$T>T_{fs}(N)$ \cite{jaklic00} mostly reducing to an entropy requirement $s(T) > s_{min}(N)$.
For the largest TL cluster with $N=36$, we estimate
$s_{min} \sim 0.07$. In frustrated systems, this restriction comes into play only at very low $T \ll J$, 
in particular for  at $\alpha \ll 1$, reflected in {the} quite accurate 
reproduction of the remanent gs entropy $s_0$. {Conversely, long-range correlations 
remain more sensitive to $N$ as revealed in $S_{{\bf q}_0}(T \to 0)$.}   

To elucidate finite-size  effects on thermodynamic quantities, we present a direct comparison of the results
 for entropy $s(T)$ 
on various $N =18 - 36$ and two different $\alpha =0.1, 0.5$ in Fig.~\ref{figfs}, both for TL and KL.
Deviations are generally very small, 
with some finite-size  discrepancies (related also to different lattice shapes) even in the limit 
$\alpha \to 0$ where exact result for TL is known to be $s_0 = 0.323$ \cite{wannier50,moessner00},
while our finite-size result mildly deviate, e.g.,  in Fig.~\ref{fig1t}a we show $s_0 =0.345$ 
(and corresponding $R_0$) obtained on $N=36$. A few conclusions directly follow: (a)
finite-size effects on thermodynamic quantities  are more pronounced for larger $\alpha \gtrsim 0.5$, both for TL and  KL, 
which can be understood in terms of larger and $N$-dependent gaps, (b) finite-size  effects are 
more visible for TL (also persisting to higher $T$), while being very small for KL. 
This has been realized already for the  isotropic $\alpha=1$ case \cite{prelovsek20}, (c)
within TL and at $\alpha \ll 1$ our $T \to 0$  results can slightly deviate from exact $s_0 = 0.323$ 
(we get, e.g.,  for $N=36$ the value $s_0=0.345$ as shown in Fig.~\ref{fig1t}a) depending on actual lattices which 
are of different  shapes, but all with PBC. On the other hand, such
deviations are apparently quite negligible within KL as the finite systems 
reproduce the known exact $s_0=0.502$. 

\section{Triangular lattice}

\subsection{Spin structure factor}

\begin{figure}[t]
	\centering
	\includegraphics[width=\columnwidth]{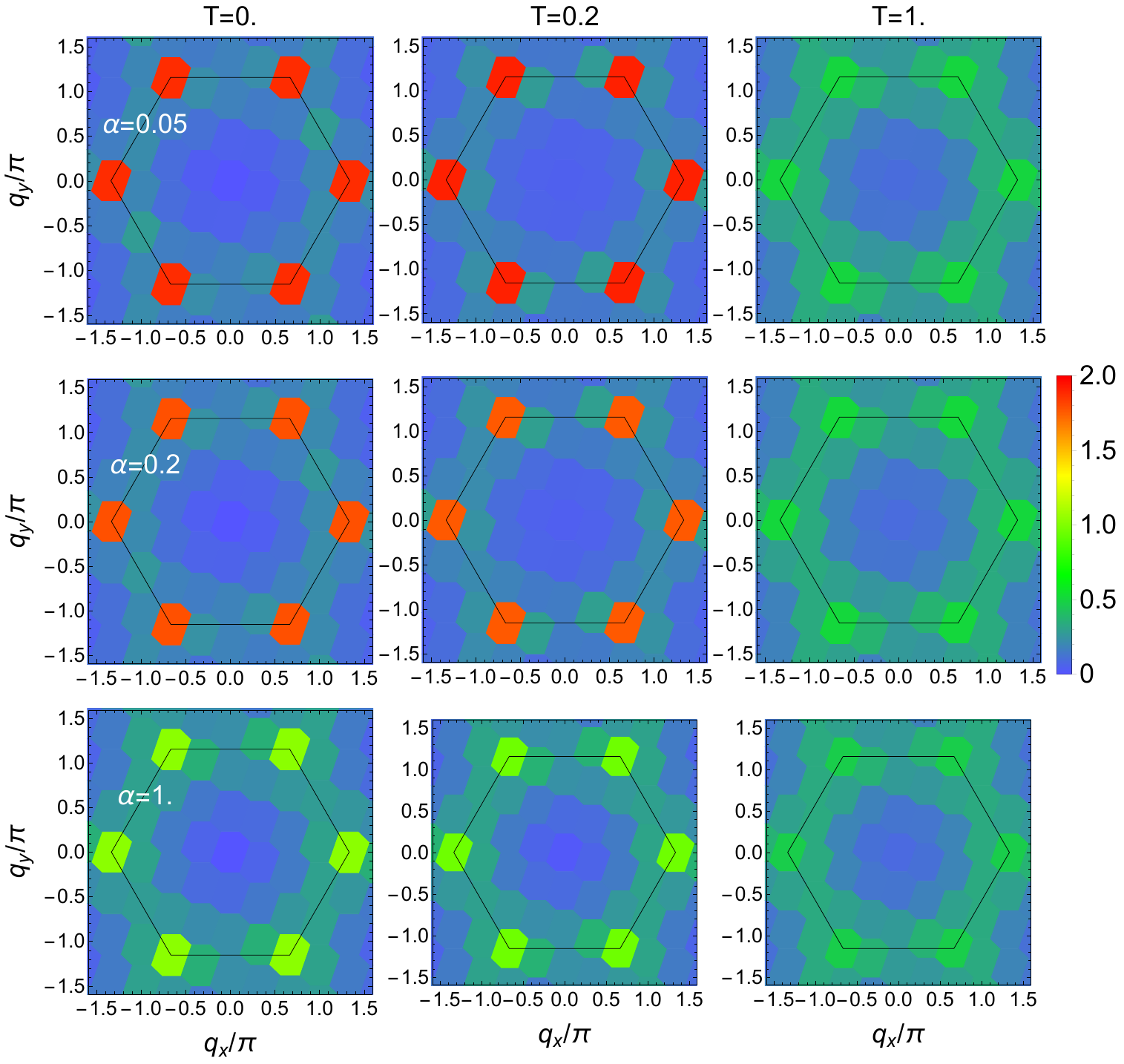}
	\caption{Color plots of the static spin structure factor $S_{\bf q}(T)$, as obtained 
	via FTLM for anisotropic HM on TL with $N=30$ sites, shown for different  $\alpha =0.05, 0.2, 1.0$ and
	{$T =0, 0.2, 1$.} }  \label{fig_grid}
\end{figure}

The spin structure factor $S_{\bf q} = (1/N) \sum_{i,j} \exp[i{\bf q}\cdot({\bf r}_i-{\bf r}_j)] S^z_i S^z_j$ 
is expected to reveal the persistence of  gs long-range spin correlations 
\cite{kleine92,kleine921,wang09,jiang09,yamamoto14,sellmann15} in the whole $\alpha \leq 1$ regime. 
Besides $T =0$ gs properties the behavior of $S_{\bf q}(T>0)$ {is} much less explored, except for the
isotropic model \cite{morita20}. Here, we present results for the $S_{\bf q}(T)$ within the anisotropic 
HM on TL, as obtained within FTLM on $N=30$ sites. In Fig.~\ref{fig_grid} we present numerical results
{for $S_{\bf q}(T)$} throughout the Brillouin zone (BZ) for ${\bf q}$ consistent with the
finite-size $N=30$  lattice with PBC, for several $\alpha =0.05, 0.2, 1.0$ and different $T =0, 0.1, 0.5$. 
Apparently, the behavior at all considered $\alpha$ is {qualitatively} similar.
At low $T${, the} results reveal very pronounced maxima at the corners of the BZ
${\bf q}_0 = (4 \pi/3, 0) $, being the signature of the LRO. It is significant that absolute and relative 
(to neighboring ${\bf q} \ne  {\bf q}_0 ${, e.g., ${\bf q_M}$ at the middle BZ edge}) maxima at ${\bf q}_0$ are even stronger in the Ising 
regime $\alpha \ll1$.  {Clearly, at $T \sim 1$ the dependence on $\alpha$ is largely washed out.}

\begin{figure}[t]
	\centering
	\includegraphics[width=0.9\columnwidth]{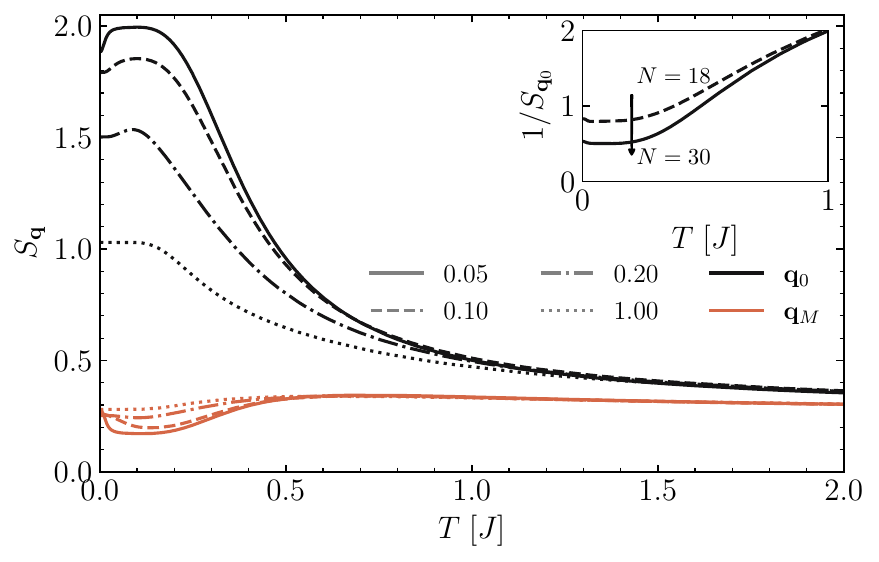}
	\caption{{The spin} structure factor $S_{\bf q}$ vs. $T$, calculated for the HM on TL with $N=30$ sites,
	{presented for ${\bf q}={\bf q}_0$  {at the BZ corner} corresponding to ordering as well as 
	 for ${\bf q}={\bf q}_M$ in the middle of the BZ edge. The results correspond to
  various anisotropies, namely $\alpha=0.05, 0.2, 0.5, 1$. The inset shows} the comparison of $S^{-1}_{{\bf q}_0}(T)$
	 for {$\alpha =0.05$} as calculated on $N=18,30$ sites, respectively.	 }    \label{fig_sqt}
\end{figure}

More detailed results on  $T$ dependence of $S_{\bf q}$  are shown  in Fig.~\ref{fig_sqt}
for the ordering ${\bf q} = {\bf q}_0$ and for more general ${\bf q}={\bf q}_M$
{in the middle of the BZ edge}. It should be noted that here we present results (for finite system $N=30$) 
in the whole range $T\geq 0$, although it is evident that  results for $S_{\bf q}(T \sim 0)$ are 
size-dependent \cite{jiang09}, due to long-range spin correlations, in particular for ${\bf q} = {\bf q}_0$. 
This {is confirmed by the comparison to} our results on $N=18$, presented in the inset of  Fig.~\ref{fig_sqt}
for $\alpha =0.05$ and ${\bf q} = {\bf q}_0$.
As expected $S_{{\bf q}_0}(T\sim 0) \propto \mu_z^2 N $ is
consistent with the gs LRO with the finite moment $\mu_z$. It is remarkable that  the fall-off of  
$S_{{\bf q}_0}$ with $T$ is quite independent of $\alpha$ and does not appear to be related to the 
typical temperatures
visible in thermodynamic quantities $s(T)$ and $\chi_0(T)$. On the other hand, as shown in Fig.~\ref{fig_sqt}
for other ${\bf q} = {\bf q}_M$ inside the BZ, our results reveal some  anomalies 
at low $T$  in the Ising regime, which seem to indicate the relation to 
$T^*  \propto \alpha $ {observed} in, e.g., $s(T)$, although we cannot exclude that they
disappear for increasing $N \to \infty$.  

\subsection{Thermodynamic quantities}

\begin{figure}[t]
	\centering
	\includegraphics[width=\columnwidth]{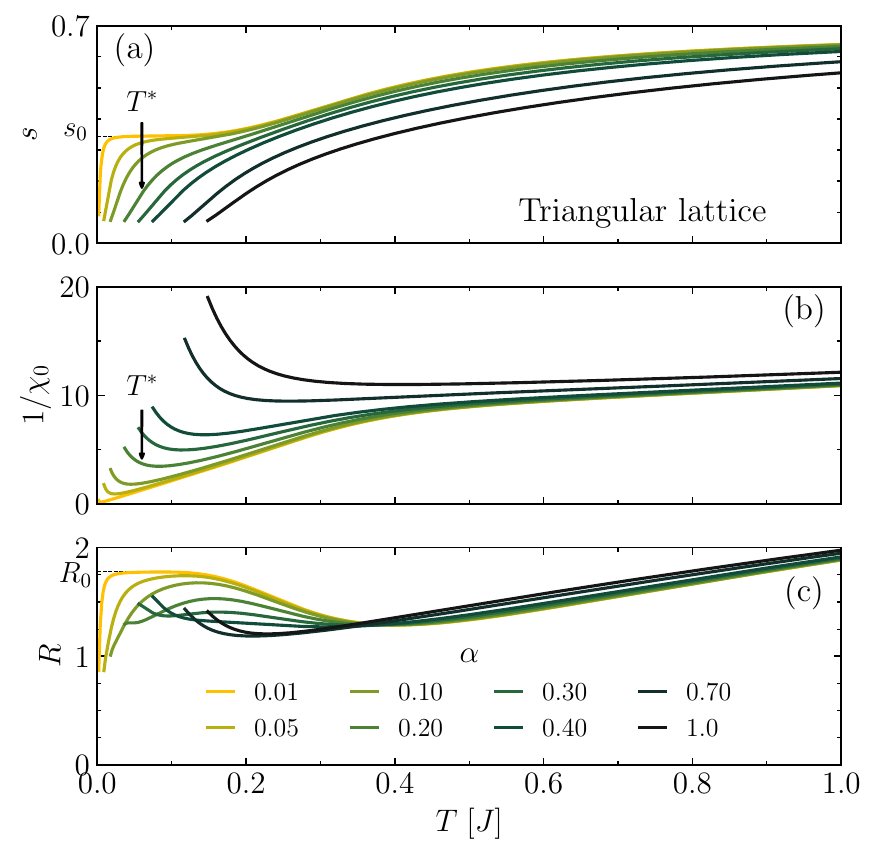}
	\caption{Entropy density $s(T)$ (a), inverse susceptibility
		$1/\chi_0(T)$ (b), and related 
		Wilson ratio $R(T)$ (c) for the Heisenberg model, as obtained with FTLM on $N=36$ TL
		for anisotropies $0< \alpha \leq 1$. 
		{Thin dashed lines mark are the residual entropy $s_0$ in the Ising limit and the corresponding 
		Wilson ratio $R_0$ while the arrows denote the crossover $T^* = 0.3 \alpha {J}$ for 
		selected $\alpha = 0.2$.}}  \label{fig1t}
\end{figure}

We present results for the  anisotropic HM on TL for various $\alpha$
between the Ising ($\alpha =0 $) and the isotropic limit ($\alpha =1$)
in Fig.~\ref{fig1t}: for the entropy density $s(T)$, inverse susceptibility 
$1/\chi_0(T)$ and the corresponding Wilson ratio $R(T)$ given by Eq.~(\ref{rw}).
All presented results in Fig.~\ref{fig1t} are restricted to estimated $s>s_{min}$ 
since below they can be dominated by various finite-size effects. 
Results in Fig.~\ref{fig1t}(a) reproduce the residual entropy $s_0$ at $\alpha \to 0$ and $T \to 0$, whereas 
the effect of $\alpha >0$ is the final
drop $s(T \ll T^*) \to 0$, where $T^* \sim 0.3  \alpha {J}$ is a characteristic crossover temperature.  
There is an evident high-$T$ regime, $T>T_0 \sim 0.4\,{J}$, where $s(T)$, as well as other quantities, 
remain weakly dependent on $\alpha$.

\begin{figure}[b]
	\centering
	\includegraphics[width=0.9\columnwidth]{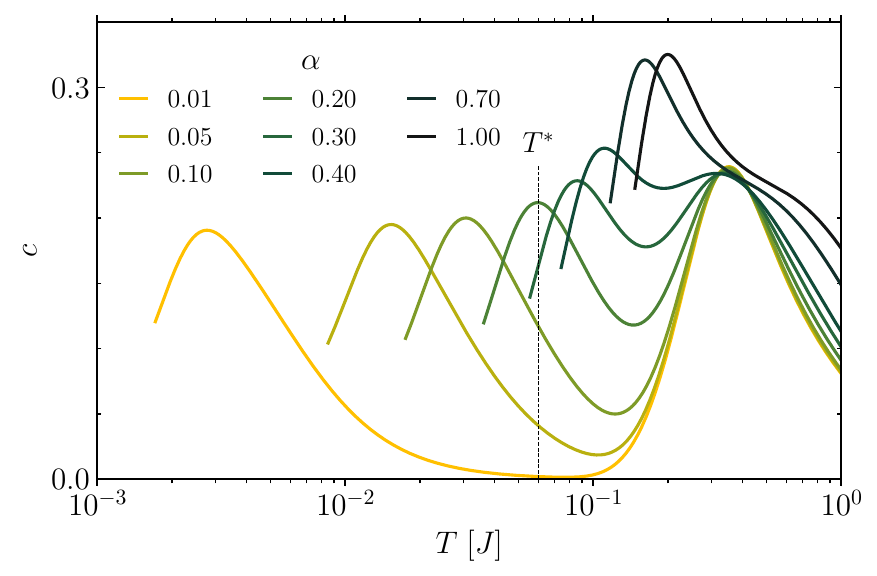}
	\caption{Specific heat $c$ vs. $T$ for the HM on TL for different $\alpha$.
		Marked is the maximum of the low-$T$ peak at $T^* =0.3 \alpha {J}$
		for $\alpha =0.2$.}   \label{fig2}
\end{figure}

{The} susceptibility $1/\chi_0(T)$ in Fig.~\ref{fig1t}(b) reveals several regimes.
For $T > T_0$ the behavior (for all $\alpha$) follows the Curie-Weiss behavior
with $\chi_0(T) \propto 1/(T+\Theta) $ where $\Theta \sim 1.5\,J$. On the other hand, in the Ising
limit ($\alpha=0$){,} the dependence turns into a Curie law $\chi_0(T < T_0) = C/T$ with $C=0.045$, 
where our value  is comparable with $C=(5/36)/4=0.035$ from  {Ref.~\onlinecite{sykes61} and with  
$C=0.042$ from Ref.~\onlinecite{sano87}. In Appendix A we present an analytical analysis that gives a
simple and quite accurate value of obtained Curie constant $C$.}

The effect of finite $\alpha >0$ is the vanishing of $\chi_0(T \to 0)=0$, leading to pronounced maximum 
at $\chi_0(T \sim T^*)$, i.e., the  minimum of $\chi_0^{-1} (T \sim T^*)$
in Fig.~\ref{fig1t}(b). The most important implication for the gs, however, follows from $R(T)$ shown in 
Fig.~\ref{fig1t}(c).  The isotropic case of $\alpha=1$ has a minimum  $R({T \sim 0.2\, J}) $ 
\cite{prelovsek20} and $R(T \to 0)$ is expected to diverge (in the thermodynamic limit) due to
the onset of magnetic LRO at $T=0$ ({note that 
$T_{fs} >0.15\,J$ is the most restrictive for $\alpha  \sim 1$}). Results shown in Fig.~\ref{fig1t}(c) indicate 
that this minimum disappears for $\alpha < \alpha^* \sim 0.3$ and the behavior changes into the 
vanishing $R(T \to 0)=0$. Approaching $\alpha \to 0$ 
a broad plateau at the Ising value $R_0 \sim  4 \pi^2C/(3s_0)$ also becomes evident and 
a downturn in $R(T)$ only occurs at $T <T^*$.  
Relevant for experiments is also the specific heat $c(T)$ presented in Fig.~\ref{fig2}, directly related to 
$s(T)$ in Fig.~\ref{fig1t}(a). Its characteristic feature is a double-peak structure, becoming very pronounced 
for $\alpha \lesssim \alpha^*$.  The high-$T$ peak at $T \sim 0.3\,{J}$ reflects correlations due to the dominant 
exchange $J$ and is nearly $\alpha${-}independent. 
On the other hand, the maximum of the lower-energy peak coincides with the 
drop of $s(T)$ in Fig.~\ref{fig1t}(a) and occurs at $T \sim T^* $.

\subsection{Lowest excitations}

\begin{figure}[t]
	\centering
	\includegraphics[width=0.9\columnwidth]{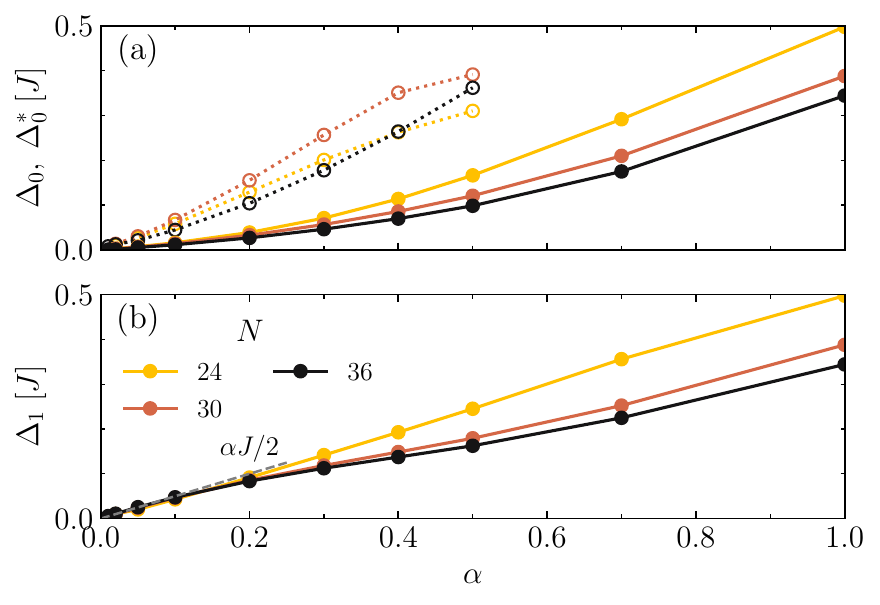}
	\caption{Magnetic and nonmagnetic gaps $\Delta_1,\Delta_0$, respectively, vs. $\alpha$, 
		as obtained on TL systems with $N=24-36$ sites. For $\alpha < 0.5$  
		next-lowest-lying nonmagnetic excitations $\Delta^*_0$ are also presented. The dashed line in the lower panel shows 
		the linear scaling of the magnetic gap on $\alpha$ in the Ising regime. }  \label{fig3}
\end{figure}

In order to understand the thermodynamic quantities, it is informative to follow the lowest excitations
within the model. Their general structure within TL
for $\alpha \leq 1$ is presented in Fig.~\ref{fig3}. The gs (at $h=0$) belongs to the nonmagnetic 
$S^z=0$ sector. In the whole $\alpha <1$ range the lowest gap $\Delta_0$ belongs to a single
nonmagnetic ($S^z=0$) state, lying below the first magnetic $S^z=1$
excitation with the gap $\Delta_1$. The next nonmagnetic gap is, however, $\Delta_0^* > \Delta_1$.
The $N$ and $\alpha$ variations of gaps are very different in $\alpha \ll 1 $ and $\alpha \sim 1$  regimes.  
In the latter, the magnetic  $\Delta_1$ is expected to vanish with increasing $N$ as $\Delta_1 \propto N^{-1}$, 
as established for  $\alpha \sim 1$ \cite{capriotti99}. This is consistent with our results in 
Fig.~\ref{fig3}.  We note that at least at $\alpha=1$, $\Delta_0$ should merge
with $\Delta_1$,  representing in this case the triplet excitation.  On the other hand, the behavior for 
$\alpha < \alpha^*$ is markedly different. Results in Fig.~\ref{fig3}
indicate that the magnetic $\Delta_1$ is almost $N$-independent and 
seems to converge to   $ \Delta_1 \sim 0.5 \alpha$. The lowest nonmagnetic
$\Delta_0 \ll \Delta_1$ {that qualitatively explains the} vanishing $R(T\to 0) \to 0$ in Fig.~\ref{fig1t}c,
whereby $\Delta_0(N)$ might even vanish for $N  \to \infty$.
Still, higher nonmagnetic excitations are above the lowest magnetic one, i.e., $\Delta_0^* > \Delta_1$.
This is in marked contrast with the analogous HM on KL, characterized by numerous nonmagnetic 
excitations below the lowest magnetic excitation 
in the whole regime of $\alpha \leq 1$, well established for $\alpha=1$ 
\cite{waldtmann98,lauchli19,prelovsek20}.

\begin{figure}[h]
\centering
\includegraphics[width=0.9\columnwidth]{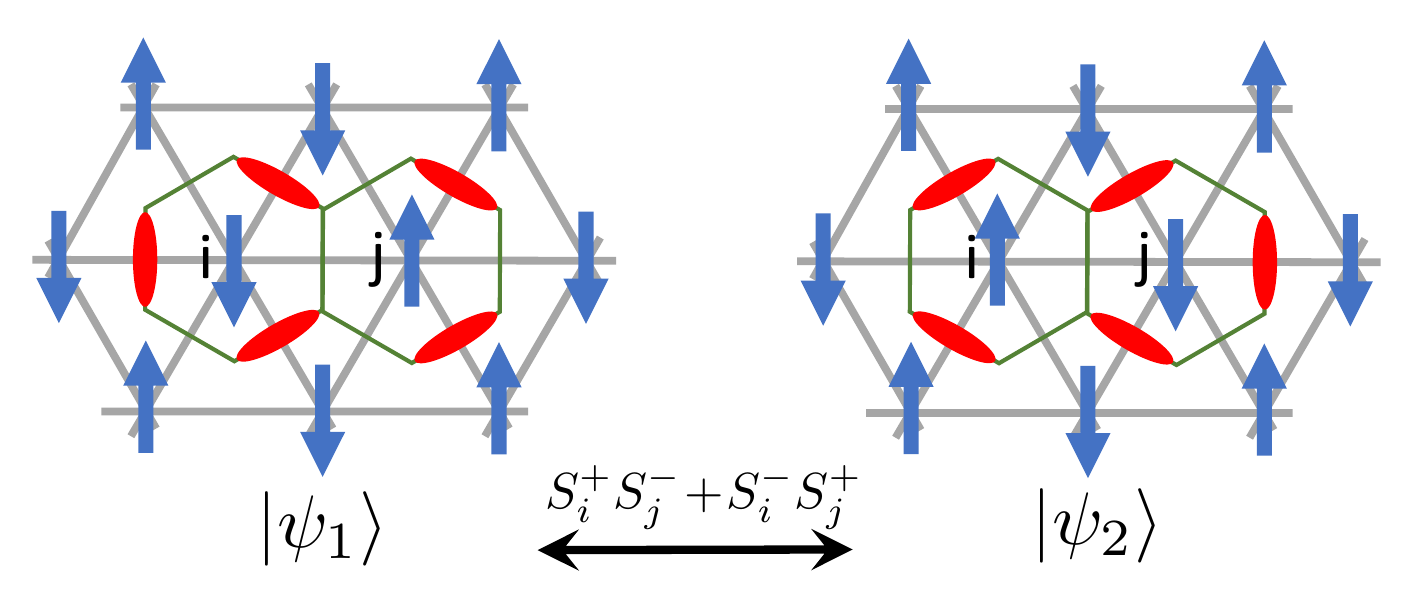}
\caption{Configurations on TL in the gs manifold which
allow a spin exchange  without changing the Ising energy. 
Green hexagons show the dual honeycomb
lattice \cite{wannier50,moessner00} with the red dimer indicating
energetically unfavorable (parallel) orientation of spins on particular TL bond.}
\label{fig_conf}
\end{figure}

The emergence of the magnetic gap $\Delta_1$ at $\alpha >0$ can be considered
through the  lifting of the Ising gs degeneracy.
For $\alpha \to 0$ one can apply the  degenerate perturbation theory, 
in analogy to the Hubbard model for large $U$ \cite{eskes94}, 
within which the  concept of ``interchangeable pairs'' of spins emerged \cite{fazekas74, kleine92}, 
treating the $\alpha$ term perturbatively within the
degenerate gs manifold.  In our case, one transforms the 
Hamiltonian in such a way that it does not change the number of
frustrated bonds. The application of the linear $\alpha $ term changes the configuration to the
one shown on the right side of Fig.~\ref{fig_conf} (denoted with
$|\psi_2\rangle$). The corresponding antisymmetric combination $|\psi_s\rangle =
(|\psi_1\rangle - |\psi_2\rangle)/\sqrt 2$ has lower energy $E_s=E_0-
\alpha J/2$ ($E_0$ is the energy of the Ising gs manifold) and $S^z=0$.
 One can also create a $S^z=1$ state
$|\psi_t\rangle=S^+_i|\psi_1\rangle$ by flipping the ``free spin'' on {the} site $i$ on the left configuration in
Fig.~\ref{fig_conf} and making spins at sites $i$ and $j$
parallel. This state has $S^z=1$ and energy $E_t=E_0$ 
(up to a linear order in $\alpha$). Within this picture follows that $\Delta_1=E_t-E_s=\alpha J/2$,
comparing favourably with FTLM results (see Fig.~\ref{fig3}) for small $\alpha$. 

\subsection{Finite fields}

\begin{figure}[t]
	\centering
	\includegraphics[width=\columnwidth]{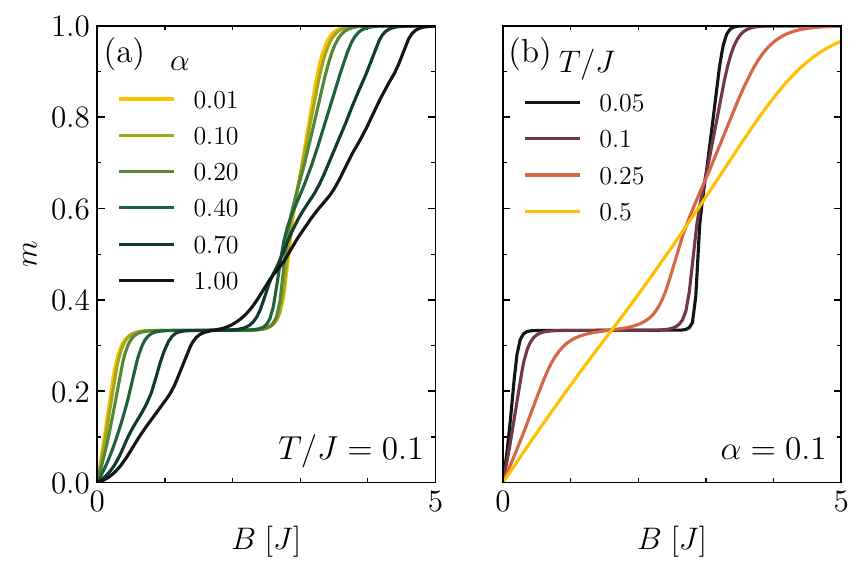}
	\caption{Magnetization curves $m(h)$ for the anisotropic HM on TL:  
		(a) for different $\alpha $ at fixed $T =0.1$,
		and (b) for different $T$  at fixed $\alpha =0.1$.}  \label{fig4}
\end{figure}

The variation of the (normalized) magnetization density $m = \langle S^z\rangle/(NS) $ with external magnetic field
$h$ in Eq.~(\ref{his}) can be evaluated within FTLM {without additional} numerical effort.
The magnetization curves $m(h)$ are of particular interest also for
the experiment since in related materials
the whole regime of $h$ can {potentially be} explored. On frustrated lattices, such as TL and KL,
a pronounced plateau at $m=1/3$ is expected and has been investigated within gs calculations
\cite{honecker04}.  The focus here is on the behavior at small finite
$\alpha \ll 1$, since in the Ising
limit ($\alpha=0$) the variation $m(h)$ is anomalous, with a discontinuous jump at $T \sim 0$,
i.e., any small $h> 0$ stabilizes the $m=1/3$ plateau. Numerical results for 
$m(h)$ for some characteristic $\alpha$  are presented  in
Fig.~\ref{fig4} where we show results up to $\alpha=1$ for completeness.  The variation with $\alpha$
at small finite $T =0.1\,{J}$ reveals that the jump at $\alpha =0$ transforms into a nearly linear
variation $m \propto h$ up to the $m=1/3$ plateau.  At the same time, the plateau
melts with increasing $T$ and essentially disappears for $T > T_0=0.4{J}$ even for small 
$\alpha$, as shown in Fig.~\ref{fig4}(b).

\section{Kagome lattice}

\subsection{Spin structure factor}

\begin{figure}[t]
	\centering
	\includegraphics[width=\columnwidth]{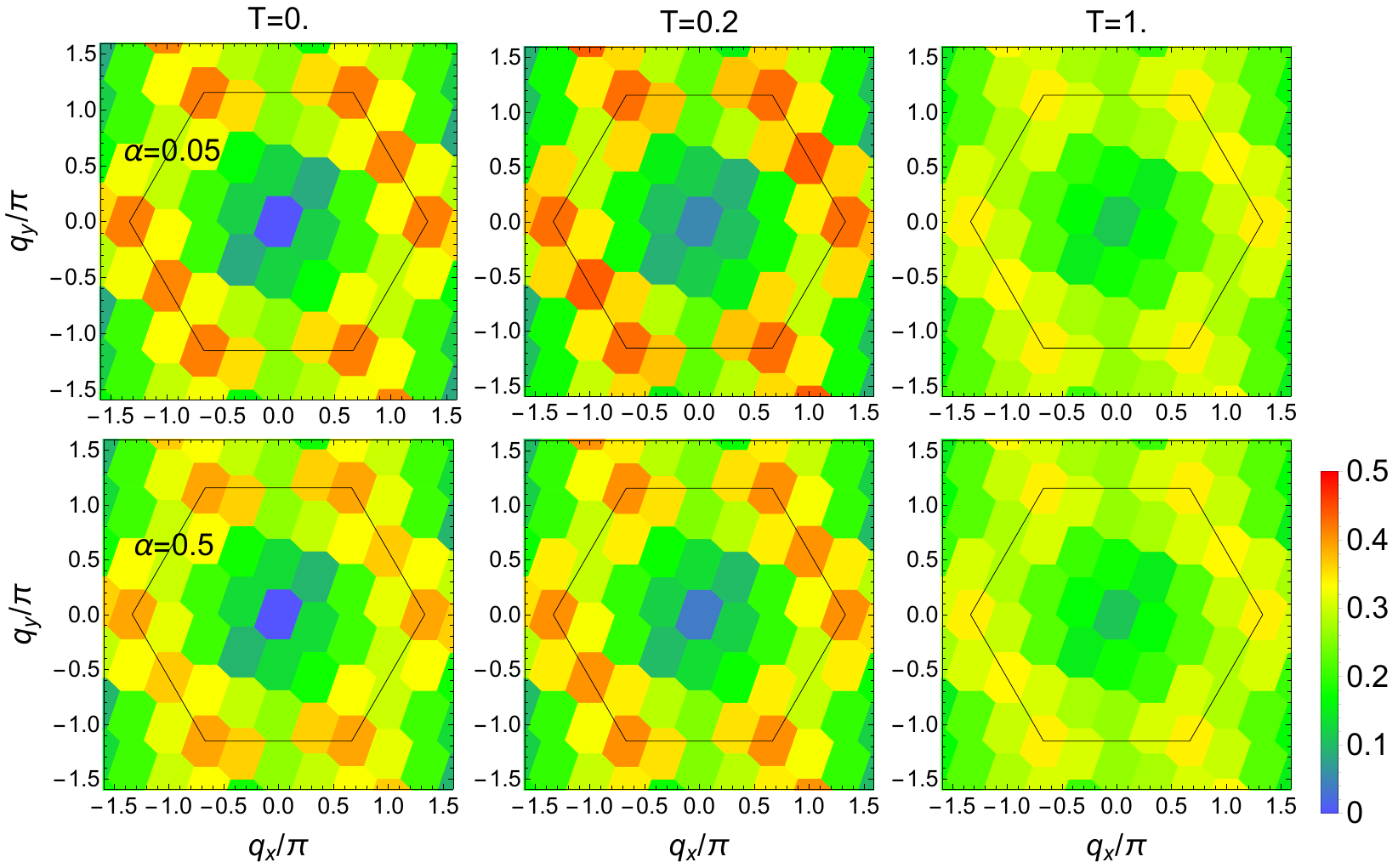}
	\caption{Color plots of the static spin structure factor $S_{\bf q}(T)$, as obtained 
	via FTLM for anisotropic HM on KL with $N=30$ sites, shown for two $\alpha =0.05, 0.5$ and
	$T =0., 0.2, 1$.}  \label{fig_gridk}
\end{figure}

In contrast to TL, gs spin correlations within the anisotropic HM on KL are expected to be {short-range}
even in the Ising limit $\alpha =0$ \cite{moessner00,moessner01}. Here, we present finite $T \geq 0$ results for 
easy-axis spin structure factor as obtained via FTLM for systems up to $N=30$ sites.
We note that for isotropic $\alpha=1$ our $S_{\bf q}(T)$ results correspond well to {previous studies}
\cite{morita20}.  In Fig.~\ref{fig_gridk} we present results in analogy with Fig.~\ref{fig_grid}, shown for 
the same $\bf q$ (taking the site/bond distance as unit $a=1$) as for TL (at same $N$). {It is quite evident 
that (in contrast to TL) the
variation of $S_{\bf q}(T)$ with ${\bf q}$ is quite smooth even in the gs with a weak maximum at the 
boundary of the extended BZ. The dependence on both $\alpha$ and $T$ is modest. This signals
very short-range spin correlations and SL character, well established in the isotropic 
$\alpha=1$ case.}
  
\subsection{Thermodynamic quantities}

\begin{figure}[h!]
\centering
\includegraphics[width=\columnwidth]{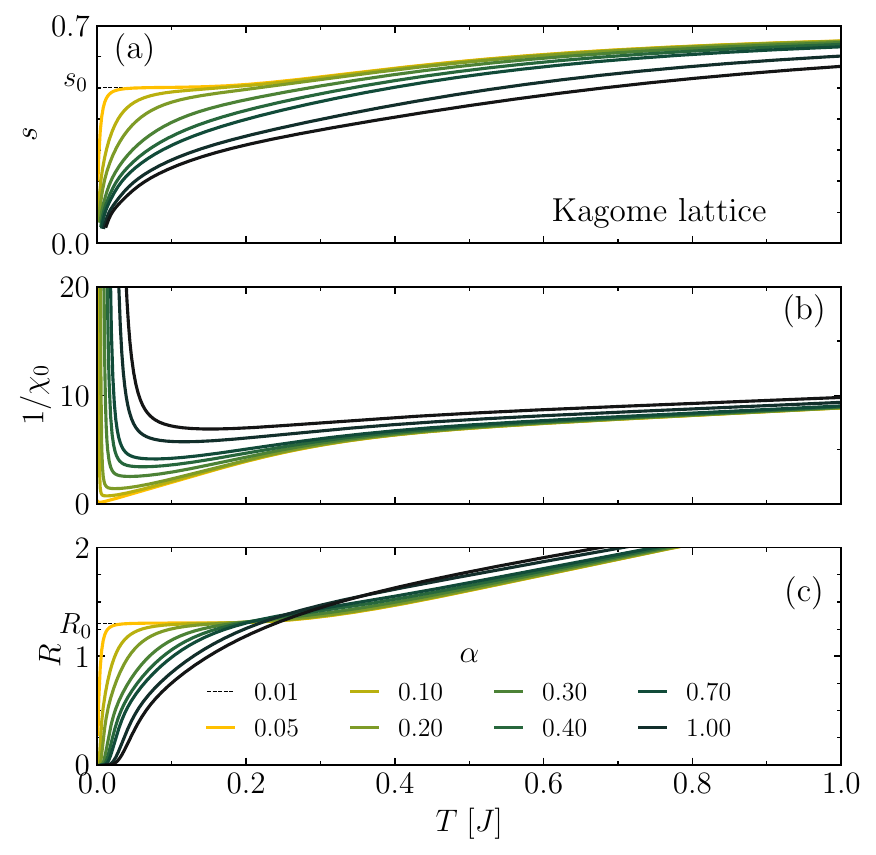}
\caption{ Thermodynamic quantities for the Heisenberg model on KL, as obtained with FTLM on 
$N=36$ sites for different $\alpha \leq 1$ : 
(a)  entropy density $s(T)$, (b) inverse susceptibility $1/\chi_0(T)$, and (c) Wilson ratio $R(T)$. 
Marked are also exact $s_0= 0.502$ as the Ising-limit result, and the corresponding 
Wilson ratio $R_0= 1.306$. }
\label{figS2}
\end{figure}

\begin{figure}[h!]
\centering
\includegraphics[width=0.9\columnwidth]{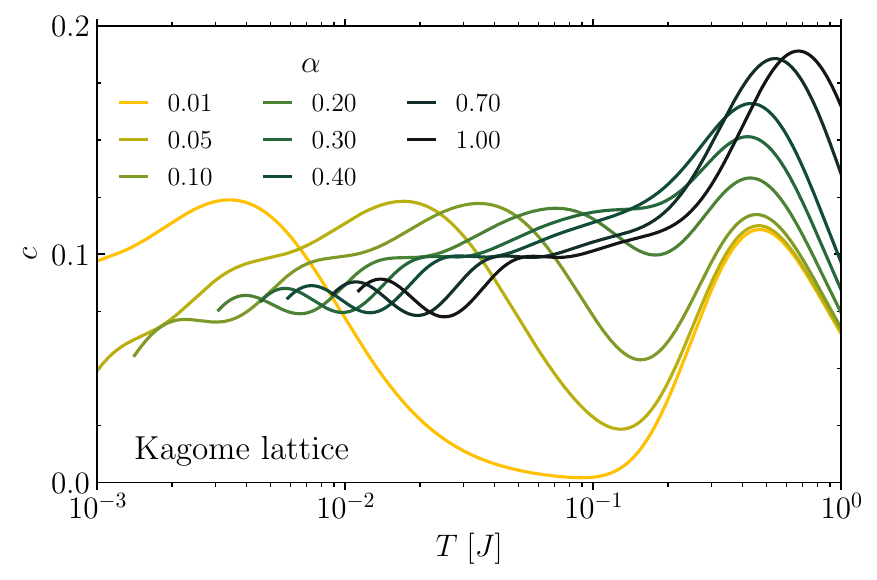}
\caption{Specific heat $c$ vs. $T$ (in log scale) for the anisotropic HM on KL
for various $\alpha$.  }  
\label{figS3}
\end{figure}

We present further results for thermodynamic quantities for the anisotropic HM on KL, 
in analogy to previous results for TL. In Fig.~\ref{figS2} results are shown for 
various $\alpha \leq 1$ for entropy density $s(T)$, inverse susceptibility $1/\chi_0(T)$ and Wilson ratio $R(T)$, as obtained 
via FTLM on the largest KL with $N=36$ sites {(the cutoff here is at $s>s_{min} =0.05$)}. 
In Fig.~\ref{figS3} the corresponding specific  heat $c(T)$ is shown.  The comparison with results on
TL in Figs.~\ref{fig1t},\ref{fig2}   reveal similarities, but also pronounced qualitative differences between 
 both lattices: 
(a) There is an essential difference close to the isotropic regime $\alpha \sim 1$, where HM on KL 
is the prominent example of a  QSL without LRO \cite{mila98,budnik04,lauchli11,iqbal13,schnack18}, 
showing up also in the smoothly vanishing $R(T \to 0)$ \cite{prelovsek18,prelovsek20}.
(b) In the regime $\alpha < \alpha^*$ for TL thermodynamic properties appear qualitatively 
similar. The drop of  $s(T)$ from the Ising value $s_0$ with the corresponding lower peak in 
$c(T)$ appears at $T \sim T^* \sim 0.5 \alpha\,{J}$.  Related is the minimum of $1/\chi_0(T)$
in Fig.~\ref{figS2}(b). (c) Still, there is a marked difference between TL and KL
in the sharpness of the lower peak in $c(T)$. As evident in Fig.~\ref{figS3} the latter peak
in KL extends to much lower $T$, which can be attributed to a large
density of low-lying nonmagnetic excitations, valid  also for the isotropic case HM at $\alpha =1$ 
\cite{waldtmann98,lauchli19,prelovsek20}.   
Additional structure apparent in $c(T)$ at lowest $T \gtrsim T_{fs}$ can  be partly attributed to finite-size
effects, as also observed for $\alpha=1$ for even larger $N=42$ \cite{schnack18}. 

\subsection{Lowest excitations}

\begin{figure}[h!]
\centering
\includegraphics[width=0.9\columnwidth]{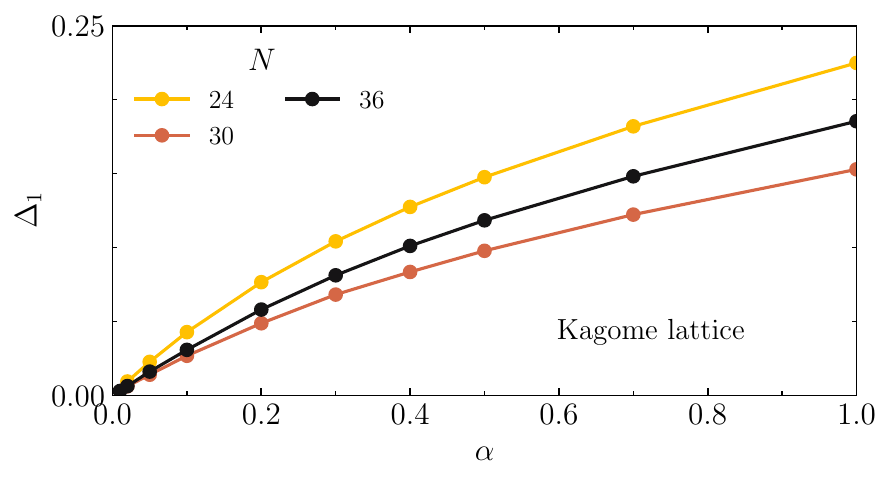}
\caption{ The magnetic gap $\Delta_1$ vs. $\alpha$, obtained on KL systems 
with $N=24-36$ sites.}  
\label{figS4}
\end{figure}

In analogy to TL, we also analyze the gap structure on KL. 
In Fig.~\ref{figS4} we present the  variation of the magnetic gap $\Delta_1$ with $\alpha$ 
for different system sizes $N$. The gap vanishes (linearly for all $N$) approaching Ising limit $\alpha \to 0$,
in analogy to TL in Fig.~\ref{fig4}. However, the gap for KL
increases steadily up to  $\alpha \lesssim 1$, which is in contrast to TL. 
The $N$ dependence is less systematic
even at $\alpha \lesssim 1$  in  accordance with the open question whether $\Delta_1$ remains {finite} 
in the  $N \to \infty$ limit \cite{waldtmann98}.
The same question applies to our results in Fig.~\ref{figS4} for the regime of $\alpha \ll1 $,
where we do not observe clear convergence with $N$, unlike the TL case in Fig.~\ref{fig3}.
However, the crucial difference to TL is the behavior {of} nonmagnetic excitations. 
It is known that in the isotropic case, there are {(macroscopically) numerous 
nonmagnetic} excitations below the lowest  magnetic one \cite{waldtmann98,lauchli19}.
Our results reveal that this remains the case in the whole regime of $\alpha \le 1$, i.e.,
we find  many $S^z=0$ states satisfying  $\Delta_0 \ll \Delta_1$, which are hard to
enumerate fully within our Lanczos-based method. 

Presented results for the HM on KL offer an important insight into the well-established
QSL state {in that its properties in the isotropic $\alpha =1 $ model are smoothly connected to 
the Ising-like regime at $\alpha \ll 1$. This contrasts with the 
corresponding HM on TL.} 
 
\subsection{Finite fields}

\begin{figure}[h!]
\centering
\includegraphics[width=\columnwidth]{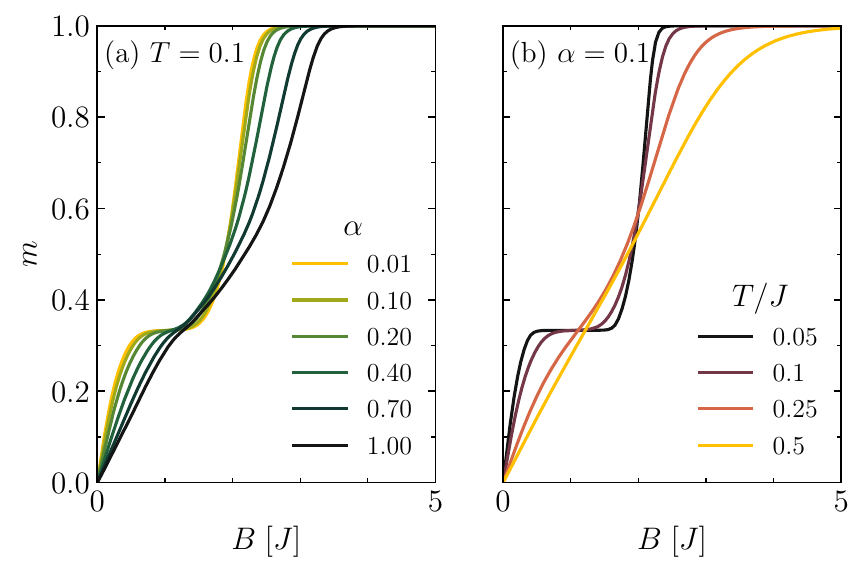}
\caption{Magnetization curves $m(h)$ for the anisotropic HM on KL:  
(a) for different $\alpha $ at fixed $T =0.1$, and (b) for different $T$  at fixed $\alpha =0.1$. }
\label{figS5}
\end{figure}

Finally, we show results for the magnetization curves $m(h)$ for KL.
Again, in the Ising limit $\alpha=0$ the variation $m(h)$ reveals a discontinuous jump at $T \sim 0$,
i.e., even small $h> 0$ stabilizes $m=1/3$ magnetization. Numerical results for 
$m(h)$ for some characteristic cases  are presented  in Fig.~\ref{figS5}.  The variation with $\alpha$
at small finite $T =0.1\,{J}$ shows that the jump at $\alpha =0$ transforms into a nearly linear
variation $m \propto h$ up to the $m=1/3$ plateau.  At the same time, the plateau
disappears with increasing $T > T_0$ already at small $\alpha \ll 1$, as shown in Fig.~\ref{figS5}(b).

\section{Discussion}	

Isotropic AFM spin models on frustrated lattices have been intensively studied, mostly
as candidates for the QSL phenomenon. The anisotropy studied here offers another
route to interesting collective phenomena. Our analysis indicates that with the 
increasing easy-axis anisotropy, the thermodynamic quantities within Heisenberg model on TL undergo
a crossover from the isotropic-like regime to the Ising regime at $\alpha < \alpha^*  \sim 0.3$, 
most pronounced in the behavior of the  Wilson ratio  $R(T \to  0)$ vanishing at $\alpha < \alpha^*$ and increasing for 
$\alpha > \alpha^*$, at least within the range of low $T$ but above finite-size $T >T_{fs}(N)$. On
the other hand, spin correlations as displayed in $S_{\bf q}(T \to 0)$ are 
consistent with LRO in the gs {in the whole $\alpha \le 1$ regime}, thus apparently coexisting with strongly $\alpha$-dependent thermodynamic
properties. It is quite remarkable that {the} calculated thermodynamic quantities, at least in the Ising regime $\alpha \ll 1$,
do not exhibit any significant finite-size effects down to { the lowest $T < T_{fs} \ll \alpha J$ 
while the gs static spin structure factor $S_{{\bf q}_0}$ 
remains consistent with gs LRO} and consequently also with finite-size ($N$) dependence  
$S_{{\bf q}_0}(T \to 0) \propto N$,  but at the same time not reflecting any evident influence of the 
quantum-fluctuation scale $T^* \propto \alpha$. 

Remarkably, in the Ising limit ($\alpha=0$){,} there are analogies between the low-$T$
thermodynamic properties  of spin models on the TL and KL. In particular{, the} existence
of remanent entropy $s_0>0$ and the Curie susceptibility $\chi_0 \sim C/T$. However, in contrast to the 
TL case, in the KL case  there is  a continuous (smooth) variation of all quantities from $\alpha \gtrsim 0$ 
regime to the most studied  isotropic $\alpha =1$ QSL. Moreover, {on} KL, contrary to TL,
there are numerous nonmagnetic excitations below the lowest magnetic one (i.e.,
the triplet at $\alpha =1$ \cite{waldtmann98,lauchli19,prelovsek20}) within the whole range of $\alpha \leq 1 $.
Still{,} there are evident differences in the 
spin correlations. In contrast to TL, within KL spin structure factor {$S_{\bf q}(T)$ smoothly varies
with ${\bf q}$ within the BZ, but only weakly depends on $T$ and $\alpha$, consistent with the 
short-range correlations and the QSL character.}

Finally, let us return to {the} potential relevance of our study for experimental realizations of anisotropic HM on TL and KL. {Recently, the TL antiferromagnet NdTa$_7$O$_{19}$ was shown to host dominant Ising spin correlations between nearest 
neighbors} and the anisotropy was estimated to be $\alpha = 0.18$ \cite{arh22}. This estimate was based on the assumption that the exchange anisotropy in the lowest order follows the anisotropy of the $g$ factor squared \cite{abragam70}. Various
experiments suggest QSL gs arising from strong Ising anisotropy of the exchange interactions. A direct comparison to 
our results is at present limited, as  susceptibility data are so far restricted to  {powder} samples at $T \gtrsim J$, and the specific heat has not 
been measured yet. Recently, the delafossite compound KTmSe$_2$ has been also proposed as another quantum-Ising 
TL candidate \cite{zheng23}.  

\section*{Acknowledgments.} 
We thank Takami Tohyama, Katsuhiro Morita, and Fr\'ed\'eric Mila for stimulating discussions.
This work is supported by the program P1-0044 and P1-0125 of the Slovenian Research Agency.
AZ acknowledges additional support by the Agency through Projects
No.~N1-0148 and No.~J1-2461. AW acknowledges support from the DFG through the Emmy Noether programme (WI 5899/1-1).

\appendix

\section{Origin of the Curie susceptibility}

\begin{figure}[h]
	\centering
	\includegraphics[width=0.9\columnwidth]{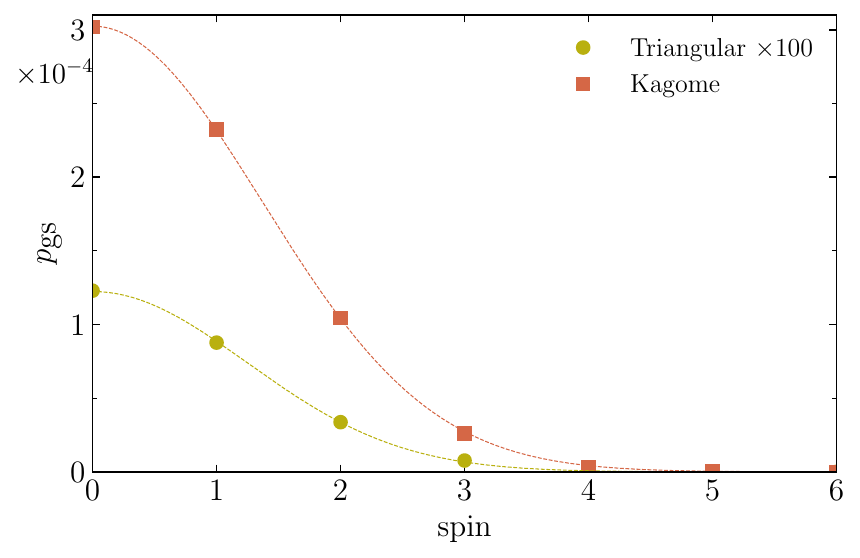}
	\caption{Probability of states with given $S^z$ in the Ising gs manifold, relative to the total number of states.
		The distribution is numerically calculated for TL and for KL on $N=36$ sites and fitted with the Gaussian (dashed lines). 
		Note that the normalized probabilities are small due to a large portion of non-free
		spins in the system.}
	\label{fig_s_gauss}
\end{figure}

In the Ising limit $\alpha=0$ the Curie susceptibility is related to
``free spins'' or ``orphans'' \cite{wannier50, isoda08, moessner00}, which
can be flipped without any energy cost within the gs manifold. From the
magnetization curves in Fig.~\ref{fig4}(a) and gs results showing
$m=1/3$  plateau one can estimate the density of free spins as $p_\textrm{free}=1/6$, 
based on the observation that any $h \gtrsim 0$ at $T=0$ leads to
$m=1/3$.   The resulting $C=p_\textrm{free}/4=0.042$ compares well with FTLM numerical
results of $C=0.045$, as obtained from Fig.~\ref{fig1t}(b).
Further support for this interpretation can be made by counting the number of states with a certain total $S^z$,
within the gs manifold. Such distribution is a Gaussian and our numerical results comply well with that (see Fig.~\ref{fig_s_gauss}).
The width of the distribution is directly related to the number of free spins
and by fitting it we get $p_\textrm{free}=0.176$, leading to the estimate $C=0.044$, 
which agrees even better with the FTLM result.

The Ising limit ($\alpha=0$) has a macroscopically degenerate gs. 
In such a case, the spin susceptibility can be written as
\begin{equation}
	\chi_0=\frac{1}{N T }\sum_{S^z} p_{S^z} (S^z)^2,
\end{equation}
where  $p_{S^z}=N_{S^z}/N_{all}$ with $N_{S^z}$ is
the number of many-body states with some value of $S^z$ and
$N_{all}$ is the total number of all states in the gs manifold.
Assuming $N_f$ free spins, each state can have a certain number
of up spins $N_\uparrow$ and down spins $N_\downarrow$ so that
$N_f=N_\uparrow+ N_\downarrow$. Further one can write the probability
for $ S^z=\frac{1}{2}(N_\uparrow - N_\downarrow)=N_\uparrow -
\frac{1}{2} N_f$ as
\begin{equation}
	p_{S^z}= \frac{1}{2^{N_f}}\binom{N_f}{N_\uparrow}\approx \sqrt{\frac{2}{\pi N_f}}\textrm{e}^{-2 (S^z)^2/N_f}~
\end{equation}
by using the normal approximation for large $N_f$ and $N_\uparrow$.
The probability of free spins becomes Gaussian for large systems and
we clearly observe such behavior numerically on $N=36$ sites within an Ising gs
manifold by counting the number of states (see
Fig.~\ref{fig_s_gauss}). Further, the fitted width of the Gaussian
is an estimate of the number of free spins $N_f$, which
gives a good estimate for the Curie constant $C=N_f/(4N)=0.044$ for TL and $C=0.051$ for KL.

\end{document}